\documentclass[12pt]{article}

\usepackage[english]{babel}
\usepackage{graphicx,rotating,epsfig}
\usepackage{a4p}

\newcommand{\TeV}{\ensuremath{\mathrm{Te\kern -0.1em V}}}
\newcommand{\GeV}{\ensuremath{\mathrm{Ge\kern -0.1em V}}}
\newcommand{\MeV}{\ensuremath{\mathrm{Me\kern -0.1em V}}}
\def\GeVc2{\ensuremath{\mathrm{ Ge\kern -0.1em V }\kern -0.2em /c^2 }}

\newcommand{\MW}{\ensuremath{M_{\mathrm{ W }}}}
\newcommand{\GW}{\ensuremath{\Gamma_{\mathrm{ W }}}}

\newcommand{\RunI}{\hbox{Run-I}}
\newcommand{\RunII}{\hbox{Run-II}}

\begin{document}


\begin{center}
{\LARGE FERMI NATIONAL ACCELERATOR LABORATORY}
\end{center}

\begin{flushright}
       TEVEWWG/WZ 2005/01 \\
       FERMILAB-TM-2330-E \\  
       CDF Note 7887 \\
       D\O\ Note 4944 \\
       hep-ex/0510077 \\
       {30 October 2005}
\end{flushright}

\vskip 1cm

\begin{center}
{\Huge \bf Combination of CDF and D\O\ Results \\[3mm]
                  on the W-Boson Width}
\vskip 1cm
{\Large
The Tevatron Electroweak Working Group\footnote{
WWW access at {\tt http://tevewwg.fnal.gov}\\
The members of the TEVEWWG who contributed significantly to the
analysis described in this note are: \\
B.~Ashmanskas (ashmanskas@fnal.gov),    
S.~Eno (eno@physics.umd.edu),           
M.W.~Gr\"unewald (mwg@fnal.gov),        
Y.-K.~Kim (ykkim@fnal.gov),             
S.~Kopp (kopp@fnal.gov),                
A.~Kotwal (kotwal@fnal.gov),            
M.~Lancaster (lmarkl@fnal.gov),         
L.~Nodulman (ljn@fnal.gov),             
Q.~Xu (qitaixu@hotmail.com),            
B.~Zhou (bzhou@umich.edu),              
J.~Zhu (junjie@fnal.gov).               
}\\[2mm]
(for the CDF and D\O\ Collaborations)
}

\vskip 1cm

{\bf Abstract}

\end{center}

{
  
  The results on the direct measurements of the W-boson width, based
  on the data collected by the Tevatron experiments CDF and D\O\ at
  Fermilab during {\RunI} from 1992 to 1996 and {\RunII} since 2001
  are summarized. The combination of the published {\RunI} and
  preliminary {\RunII} results, taking correlated uncertainties
  properly into account, is presented.  The resulting preliminary
  Tevatron average for the total decay width of the W boson is:
  $\GW=2078\pm87~\MeV$, where the total error consists of a
  statistical part of $62~\MeV$ and a systematic part of $60~\MeV$.

}

\vfill



\section{Introduction}

The experiments CDF and D\O, taking data at the Tevatron
proton-antiproton collider located at the Fermi National Accelerator
Laboratory, have made several direct experimental measurements of the
total decay width $\GW$ of the W boson.  The published
measurements~\cite{GW-CDF-Ia, GW-CDF-Ib, GW-D0-I} are based on {\RunI}
data (1992-1996) while the results from {\RunII} are
preliminary~\cite{GW-D0-II-0408}.  They utilize the
electron~\cite{GW-CDF-Ia, GW-CDF-Ib, GW-D0-I, GW-D0-II-0408} and
muon~\cite{GW-CDF-Ib} decay topologies arising in W-boson decay.  The
recently presented preliminary measurement in the electron
channel~\cite{GW-D0-II-0408} by the D\O\ collaboration is based on a
large {\RunII} data set with correspondingly well controlled
systematic uncertainties.

This note reports on the combination of these measurements, using the
published {\RunI} measurements~\cite{GW-CDF-Ia, GW-CDF-Ib, GW-D0-I}
combined earlier~\cite{MWGW-RunI-PRD} and in particular including the
preliminary {\RunII} measurement from D\O~\cite{GW-D0-II-0408}.  The
combination presented here takes into account the statistical and
systematic uncertainties as well as the correlations between
systematic uncertainties, and replaces the previous
combination~\cite{MWGW-RunI-PRD}.  The new preliminary {\RunII} {D\O}
measurement~\cite{GW-D0-II-0408} is the single most precise W-boson
width measurement and has the largest weight in this new combination.

\section{Measurements}

The five measurements of $\GW$ to be combined are listed in
Table~\ref{tab:inputs}.  Besides central values and statistical
uncertainties, the systematic errors arising from various sources are
reported in Table~\ref{tab:inputs}.  For each measurement, the
individual error contributions are combined in quadrature.  The
sources of systematic errors considered here are the same as in the
previous combination~\cite{MWGW-RunI-PRD}:\footnote{The labels in
parenthesis are those used in Table~\ref{tab:inputs}.}

\begin{itemize}
\item Lepton energy scale (E-scale)
\item Lepton $E$ or $p_T$ non-linearity (Non-lin)
\item Detector response to QCD radiation and underlying event $p_T$
(Recoil)
\item Transverse momentum distribution of the W bosons ($p_T(W)$)
\item Backgrounds (BG)
\item Detector modeling and lepton ID (DM, $\ell$-ID), including
distribution of the primary vertex, calorimeter position resolution,
recoil momentum biases and efficiencies along lepton direction, and
the selection bias
\item Lepton resolution (Resol.)
\item Parton distribution functions (PDF)
\item QED radiative corrections (QED RC)
\item W-boson mass ($\MW$)
\end{itemize}
The details on these sources of systematic uncertainties are given in
Reference~\cite{MWGW-RunI-PRD} and the individual publications of the
two experiments~\cite{GW-CDF-Ia, GW-CDF-Ib, GW-D0-I, GW-D0-II-0408}.
Note that the error source called ``parton luminosity slope as a
function of $Q^2$ (PLS)'' considered in the previous
combination~\cite{MWGW-RunI-PRD} for the D\O\ Run-I
measurement~\cite{GW-D0-I} is combined in quadrature with the PDF
uncertainty.

\begin{table}[t]
\begin{center}
\renewcommand{\arraystretch}{1.30}
\begin{tabular}{|l||rrr|r||r|}
\hline       
       & \multicolumn{4}{|c||}{{\RunI}} & \multicolumn{1}{|c|}{{\RunII}} \\
\hline
       & \multicolumn{3}{|c|}{ CDF } & \multicolumn{1}{|c||}{ D\O\ } & \multicolumn{1}{|c|}{ D\O\ } \\
\hline       
       &  $e$ (Ia) &  $e$ (Ib) & $\mu$&  $e$ & $e$ \\
\hline 			       
\hline 			       
Result & 2110 & 2175 & 1780 & 2231 & 2011 \\
\hline                         
\hline                         
Stat.  &  280 &  125 &  195 &  142 &   93 \\
\hline                         
\hline                         
E-scale&   42 &   20 &   15 &   42 &   23 \\
Non-lin&    - &   60 &    5 &    - &    - \\
Recoil &  103 &   60 &   90 &   59 &   80 \\
$p_T(W)$& 127 &   55 &   70 &   12 &   29 \\
BG     &   17 &   30 &   50 &   42 &    3 \\
DM, $\ell$-ID  
       &    - &   30 &   40 &   10 &   16 \\
Resol. &   13 &   10 &   20 &   27 &   51 \\
\hline
PDF    &   15 &   15 &   15 &   39 &   27 \\
QED RC &   28 &   10 &   10 &   10 &    3 \\
$\MW$  &   10 &   10 &   10 &   15 &   15 \\
\hline                         
Syst.  &  173 &  114 &  135 &   99 &  108 \\
\hline                         
\hline                         
Total  &  329 &  169 &  237 &  173 &  142 \\
\hline
\end{tabular}
\end{center}
\caption[Input measurements]{Summary of the five measurements of $\GW$
performed by CDF and D\O. All numbers are in $\MeV$.  For each
measurement, the corresponding column lists experiment and channel,
central value and contributions to the total error, namely statistical
error and systematic errors arising from various sources defined in
the text.  Overall systematic errors and total errors are obtained by
combining individual errors in quadrature. }
\label{tab:inputs}
\end{table}

Further studies on the systematic errors are necessary to achieve
better understanding and will be pursued in the future.  The described
procedure with the quoted numbers represent our current understanding
of the various error sources and their correlations, adequate for the
precision of the current measurements and their combination presented
here.

\section{Combination}

In the combination, the error contributions arising from different
sources are assumed uncorrelated between measurements.  The
correlations of error contributions arising from the same source are
the same as in the previous combination~\cite{MWGW-RunI-PRD}:
\begin{itemize} 
\item uncorrelated between any two measurements: statistical error,
energy scale, non-linearity, recoil model, $p_T(W)$, background,
detector modeling and lepton ID, lepton resolution, parton luminosity
slope;
\item 100\% correlated between any two measurements: parton
distribution functions, QED radiative corrections, W-boson mass.
\end{itemize}
The resulting matrix of global correlation coefficients is listed in
Table~\ref{tab:correl}.

\begin{table}[h]
\begin{center}
\renewcommand{\arraystretch}{1.30}
\begin{tabular}{|lr||rrr|r||r|}
\hline       
    &  & \multicolumn{4}{|c||}{{\RunI}} & \multicolumn{1}{|c|}{{\RunII}} \\
\hline
    &  & \multicolumn{3}{|c|}{ CDF } & \multicolumn{1}{|c||}{ D\O\ } & \multicolumn{1}{|c|}{ D\O\ } \\
\hline       
    &  &  $e$ (Ia) &  $e$ (Ib) & $\mu$&  $e$ & $e$ \\
\hline 			       
\hline 			       
CDF-Ia & $e$   &  1.00&      &      &      &       \\
CDF-Ib & $e$   &  0.01&  1.00&      &      &       \\
CDF-I  & $\mu$ &  0.01&  0.01&  1.00&      &       \\
\hline
D\O-I  & $e$   &  0.02&  0.03&  0.02&  1.00 &      \\
\hline
\hline
D\O-II & $e$   &  0.01&  0.02&  0.02&  0.05 & 1.00 \\
\hline
\end{tabular}
\end{center}
\caption[Global correlations between input measurements]{Matrix of
global correlation coefficients between the measurements of $\GW$.}
\label{tab:correl}
\end{table}

The measurements are combined using a program implementing a numerical
$\chi^2$ minimization as well as the analytic BLUE
method~\cite{Lyons:1988, Valassi:2003}. The two methods used are
mathematically equivalent, and are also equivalent to the method used
in the previous combination~\cite{MWGW-RunI-PRD}, and give identical
results for the combination. In addition, the BLUE method yields the
decomposition of the error on the average in terms of the error
categories specified for the input measurements~\cite{Valassi:2003}.

\section{Results}

The combined value for the W-boson width is:
\begin{eqnarray}
\GW & = & 2078\pm87~\MeV\,,
\end{eqnarray}
where the total error of $87~\MeV$ contains the following components:
a statistical error of $62~\MeV$; and systematic error contributions
of: energy scale $14~\MeV$, non-linearity $15~\MeV$, recoil model
$36~\MeV$, $p_T(W)$ $21~\MeV$, background $14~\MeV$, detector modeling
and lepton ID $11~\MeV$, lepton resolution $19~\MeV$, parton
distribution functions $25~\MeV$, QED radiative corrections $9~\MeV$,
W-boson mass $13~\MeV$, for a total systematic error of $60~\MeV$.

\begin{figure}[p]
\begin{center}
\includegraphics[width=0.8\textwidth]{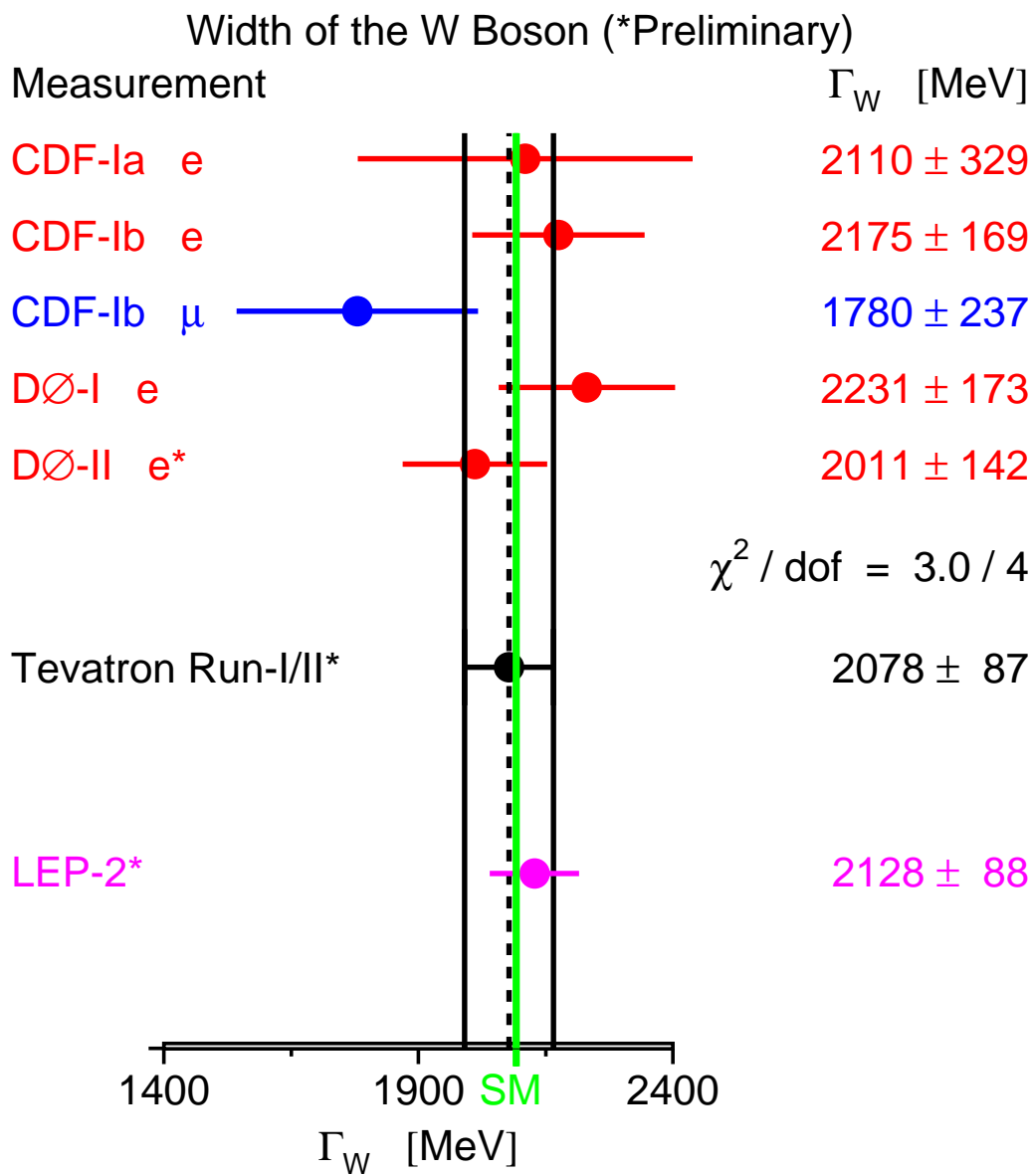}
\end{center}
\caption[Comparison of the measurements of the W-boson width]
{Comparison of the measurements of the W-boson width and their
average.  The most recent preliminary result from
LEP-2~\cite{LEPEWWG0405} and the Standard Model prediction are also
shown.}
\label{fig:gw-bar-chart} 
\end{figure}

The $\chi^2$ of this average is 2.99 for 4 degrees of freedom,
corresponding to a probability of 56\%, showing that all measurements
are in good agreement with each other which can also be seen in
Figure~\ref{fig:gw-bar-chart}.  The pull of each measurement with
respect to the average and the weight of each measurement in the
average are reported in Table~\ref{tab:stat}.

\begin{table}[h]
\begin{center}
\renewcommand{\arraystretch}{1.30}
\begin{tabular}{|l||rrr|r||r|}
\hline       
       & \multicolumn{4}{|c||}{{\RunI}} & \multicolumn{1}{|c|}{{\RunII}} \\
\hline
       & \multicolumn{3}{|c|}{ CDF } & \multicolumn{1}{|c||}{ D\O\ } & \multicolumn{1}{|c|}{ D\O\ } \\
\hline       
       &  $e$ (Ia) &  $e$ (Ib) & $\mu$&  $e$ & $e$ \\
\hline 			       
\hline 			       
Pull   & $+0.10$ & $+0.67$ & $-1.35$ & $+1.02$ &  $-0.59$ \\
\hline
Weight [\%]
       & $  6.3$ & $ 24.6$ & $ 12.4$ & $ 22.3$ &  $ 34.4$ \\
\hline
\end{tabular}
\end{center}
\caption[Pull and weight of each measurement]{Pull and weight of each
  measurement in the average.
}
\label{tab:stat} 
\end{table} 

\section{Summary}

A preliminary combination of measurements of the total decay width of
the W boson, $\GW$, from the Tevatron experiments CDF and D\O\ is
presented.  The combination includes the four published {\RunI}
measurements and one preliminary {\RunII} measurement.  Taking into
account statistical and systematic errors including their
correlations, the preliminary Tevatron result is:
$\GW=2078\pm87~\MeV$.

\clearpage

\bibliographystyle{tevewwg} 
\bibliography{gw0508}

\end{document}